\documentclass[letterpaper,aps,prl,twocolumn,superscriptaddress]{revtex4-2}
\usepackage{graphicx}% Include figure files
\usepackage{dcolumn} % Align table columns on decimal point
\usepackage{bm}      % bold math
\usepackage{amsmath}
\usepackage{xcolor}
\usepackage[colorlinks]{hyperref}
\hypersetup{
     colorlinks = true,
     linkcolor = blue,
     anchorcolor = blue,
     citecolor = blue,
     filecolor = blue,
     urlcolor = blue
     }
\newcommand{\rxx}{$\rho_{xx}$}
\newcommand{\rxy}{$\rho_{xy}$}
\newcommand{\sxx}{$S_{xx}$}
\newcommand{\sxy}{$S_{xy}$}
\newcommand{\axy}{$\alpha_{xy}$}

\begin{document}

\title{Quantized Thermoelectric Hall Plateau in the Quantum Limit of Graphite as a Nodal Line Semimetal}

\author{Andhika~Kiswandhi}
\email{kiswandhi@issp.u-tokyo.ac.jp}
\affiliation{Institute for Solid State Physics, University of Tokyo, 5-1-5 Kashiwanoha, Kashiwa, Chiba 277-8581, Japan}

\author{Tomotaka~Ochi}
\affiliation{Institute for Solid State Physics, University of Tokyo, 5-1-5 Kashiwanoha, Kashiwa, Chiba 277-8581, Japan}

\author{Toshihiro~Taen}
\affiliation{Institute for Solid State Physics, University of Tokyo, 5-1-5 Kashiwanoha, Kashiwa, Chiba 277-8581, Japan}

\author{Mitsuyuki~Sato}
\affiliation{Institute for Solid State Physics, University of Tokyo, 5-1-5 Kashiwanoha, Kashiwa, Chiba 277-8581, Japan}

\author{Kazuhito~Uchida}
\affiliation{Institute for Solid State Physics, University of Tokyo, 5-1-5 Kashiwanoha, Kashiwa, Chiba 277-8581, Japan}

\author{Toshihito Osada}
\email{osada@issp.u-tokyo.ac.jp}
\affiliation{Institute for Solid State Physics, University of Tokyo, 5-1-5 Kashiwanoha, Kashiwa, Chiba 277-8581, Japan}
\date{\today}% It is always \today, today,

\begin{abstract}
We performed thermoelectric Hall conductivity \axy~measurements on single-crystal graphite in the quantum limit up to 13 T. Both electrical and thermoelectric transport measurements were performed on the same crystal to extract pure \axy, avoiding any sample quality dependence. The \axy~converges to a plateau in the quantum limit with a linear dependence on temperature. This behavior is analogous to the quantized thermoelectric Hall effect (QTHE) observed in three-dimensional Dirac/Weyl nodal-point semimetals, and experimentally confirms a theoretical proposal on the QTHE in semimetals with nodal lines as in graphite.
\end{abstract}

%\pacs{72.80.Ga, 75.50.Pp, 71.38.-k, 62.50.-p, 61.05.cp}% PACS, the Physics and Astronomy

\maketitle
\section{I. INTRODUCTION}
Thermoelectric effects in three-dimensional (3D) Dirac/Weyl semimetals (DWSM) under a high magnetic field have been theoretically investigated in recent years. Under a dissipationless condition, the thermoelectric Hall conductivity \axy~was found to converge to a plateau, proportional to temperature ($T$), but independent of carrier density and magnetic field strength ($B$) upon entering the quantum limit (QL) \cite{Kozii}. This phenomenon is known as the quantized thermoelectric Hall effect (QTHE). Here, \axy~is an off-diagonal element of the thermoelectric conductivity tensor $\overset{\text{\tiny$\leftrightarrow$}}{\alpha}$ defined by $\textbf{j} = \overset{\text{\tiny$\leftrightarrow$}}{\sigma} \textbf{E} + \overset{\text{\tiny$\leftrightarrow$}}{\alpha} (-\boldsymbol{\nabla} T)$, where $\textbf{j}$, $\textbf{E}$, $\overset{\text{\tiny$\leftrightarrow$}}{\sigma}$, and $-\boldsymbol{\nabla}{T}$ are the current density, electric field, electrical conductivity tensor, and temperature gradient, respectively. The Seebeck coefficient \sxx~is the diagonal element of the thermopower tensor $\overset{\text{\tiny$\leftrightarrow$}}{S} = \overset{\text{\tiny$\leftrightarrow$}}{\sigma}^{-1} \overset{\text{\tiny$\leftrightarrow$}}{\alpha}$. QTHE originates from the gapless chiral $N = 0$ Landau levels (LLs) in 3D DWSM with an energy-independent density of states (DOS), which distinguish them from single-band metals. The QTHE together with the gapless chiral LLs imply \sxx~grows linearly with $B$ without an upper limit \cite{Skinner}. Such properties make 3D DWSM attractive for realizing a tunable, high-performance thermoelectricity-based power generation at low temperatures, where other materials are impractical.
\begin{figure}[htbp]
\linespread{1}
\begin{center}
\includegraphics[width=3.25in]{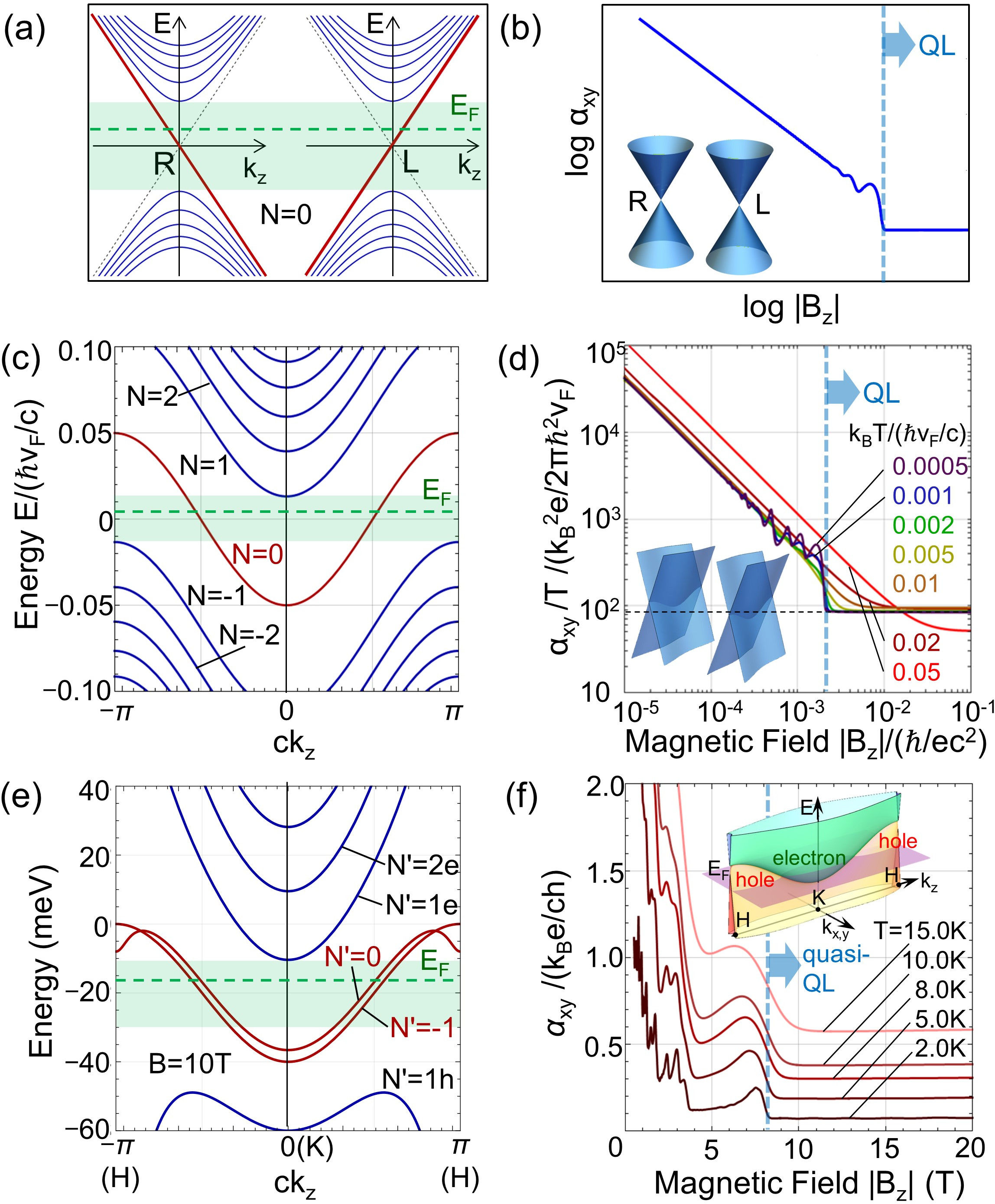}
\end{center}
\caption{(Color online) Landau subband dispersion and $B$ dependence of \axy~for (a) and (b) Dirac/Weyl nodal-point semimetals [(b) is a schematic, after \cite{Kozii}], (c) and (d) a semimetal with straight Dirac nodal-lines, and (e) and (f) bulk graphite (after Ref. \cite{Osada}). Here, $v_{\text{F}}$, $t_c$, and $c$ are the in-plane Fermi velocity, interlayer transfer integral, and interlayer spacing of multilayer semimetals with straight Dirac nodal lines, respectively. The insets of (b), (d), and (f) illustrate the band dispersion of the Dirac/Weyl nodal point semimetal, band dispersion of the semimetal with straight Dirac nodal-lines, and band dispersion of graphite, respectively. (c) was calculated using $2t_c/(\hbar v_{\text{F}}/c) = 0.05$ and $\left|B_z\right|/(\hbar/ec^2) = 0.002$, and (d) was calculated using $(n - p)c^3 = 1 \times 10^{-4}$.}\label{Fig1}
\end{figure}
Experimentally, the $B$-linear increase of \sxx~at QL has been reported in a 3D Dirac semimetal with a small spin-orbit gap Pb$_{1-x}$Sn$_{x}$Se and Weyl semimetal TaP \cite{Liang, Han}. A feature consistent with the \axy~plateau has been observed in TaP and the 3D Dirac semimetal ZrTe$_{5}$ \cite{Han, Zhang}. In ZrTe$_{5}$, although \sxx~appears to not strictly follow the $B$-linear behavior, attributed to a possible variation in the carrier balance, \axy~remains approaching a constant value at high $B$, in agreement with the theory that \axy~is independent of carrier balance. Therefore, a quantized \axy~plateau can be taken as a signature of a 3D DWSM \cite{Han, Zhang}.

However, QTHE is not necessarily unique to 3D Dirac/Weyl \textit{nodal-point semimetals}. Our simulation using a straight Dirac \textit{nodal line} semimetal model, equivalent to a stack of 2D Dirac fermion layers, yields a similar energy-independent DOS for its lowest LL. Here, we denote LLs of DWSM and nodal-line semimetal by $N$, while those of graphite by $N'$. The LL structure of the nodal-line semimetal [Fig. \ref{Fig1}(c)] shows a nonchiral $N = 0$ LL, but around the Fermi level it shows a similar configuration as the chiral $N = 0$ LLs of a pair of Dirac/Weyl cones in a 3D DWSM shown in Fig. \ref{Fig1}(a) \cite{Osada}. The calculated $B$ dependence of \axy~in the dissipationless case [Fig. \ref{Fig1}(d)] shows qualitatively the same plateau behavior at the QL as that predicted for DWSMs in Ref. \cite{Kozii} shown schematically in Fig. \ref{Fig1}(b).

Such a straight nodal line semimetal configuration can be found in graphite. The LL subband dispersion for graphite [Fig. \ref{Fig1}(e)], calculated using the Slonczewski-Weiss-McClure model \cite{Slonczewski, McClure1}, with trigonal warping ignored shows the conduction and valence bands touch along the $H–K–H$ edge in the \textbf{k} space, forming a straight nodal line \cite{Mikitik}. This configuration is very similar to that of the straight nodal-line semimetal in Fig. \ref{Fig1}(c), except that there are two lowest LLs with $N'=0, -1$, corresponding to the doubly degenerate lowest LL of the bilayer graphene stacking unit. Therefore, the QTHE can be expected in graphite at the quasi-QL where the chemical potential $\mu$ crosses only the $N' = 0, -1$ subbands as shown in Fig. 1(f) \cite{Osada}.

The electrical resistivities (longitudinal resistivity \rxx~and Hall resistivity \rxy, where $\overset{\text{\tiny$\leftrightarrow$}}{\rho} = \overset{\text{\tiny$\leftrightarrow$}}{\sigma}^{-1}$) and the Seebeck coefficient \sxx~of graphite up to the quasi-QL have been extensively studied \cite{Soule1, Soule2, Woollam}, with their Nernst coefficient \sxy~under a magnetic field explored only recently \cite{Zhu, Fauque}. However, partial measurements on separate crystals are not ideal for $\alpha_{xy}(B)$ since \rxx, \rxy, \sxx, and \sxy~may vary with different crystals. In this paper, we experimentally confirm QTHE in graphite by performing transport and thermoelectric measurements on the same graphite single crystal.

\begin{figure}[tbp]
\linespread{1}
\begin{center}
\includegraphics[width=3.25in]{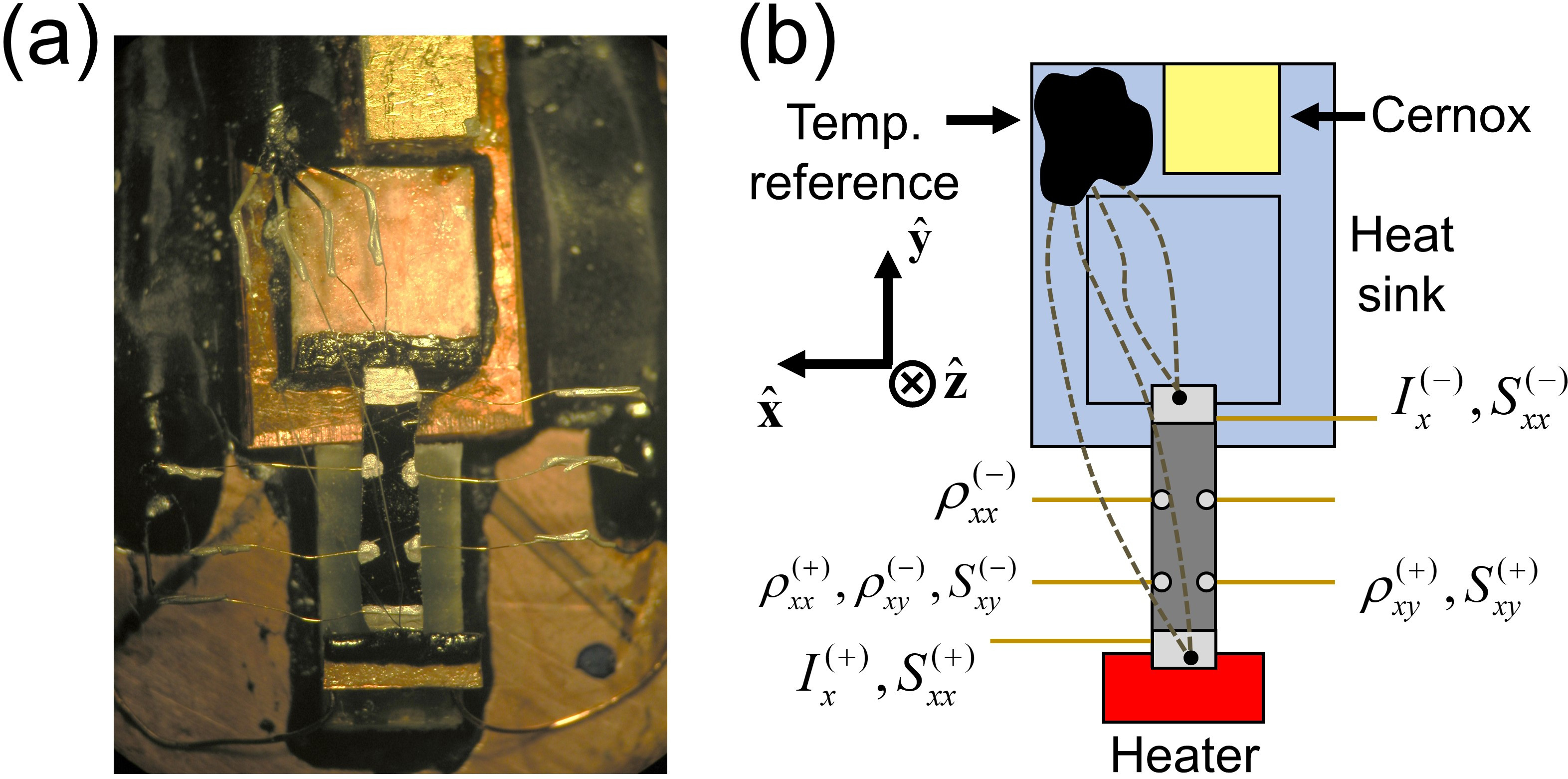}
\end{center}
\caption{(Color online) (a) Photograph of the experimental setup and (b) schematic of the experimental setup corresponding to (a). The sample is attached between a heater and copper heat sink. Two thermocouples (dashed line) measure the temperatures on the heater side and heat sink side of the sample. The thermocouples were thermally referenced to the heat sink adjacent to a Cernox thermometer. The symbols denote electrode connections to the sample, where ($+$) and ($-$) superscripts denote positive and negative electrodes, respectively.}\label{Fig2}
\end{figure}

\section{II. EXPERIMENTAL}
A bulk graphite sample with a dimension of 37$\times$0.8$\times$0.065 mm$^3$ was prepared by cleaving a Kish graphite crystal using an adhesive tape. On the clean surface, six gold wires were attached in a standard Hall configuration. The sample was placed on a thermoelectric measurement platform (Fig. \ref{Fig2}), on which electrical transport measurements were also performed. For the thermoelectric measurements, two chromel-constantan thermocouples were attached to the sample. All measurements were performed with the dc mode under a magnetic field up to 13 T parallel to the stacking direction. To eliminate mixing between \rxx (\sxx) and \rxy (\sxy) signals that may arise due to contact misalignment, signals were collected in both positive and negative field directions, then standard symmetrization and antisymmetrization procedures were used. The reported \sxx~values are relative to the gold wire electrodes with $S_{\text{Au}}\approx 1$ $\mu$V/K up to 30 T \cite{Choi1, Choi2}, so it should be understood not to deduce the carrier type using \sxx~directly. (See also the Supplemental Material for details \cite{Suppl})

\section{III. RESULTS AND DISCUSSION}
\begin{figure*}[tbp]
\linespread{1}
\begin{center}
\includegraphics[width=5.5in]{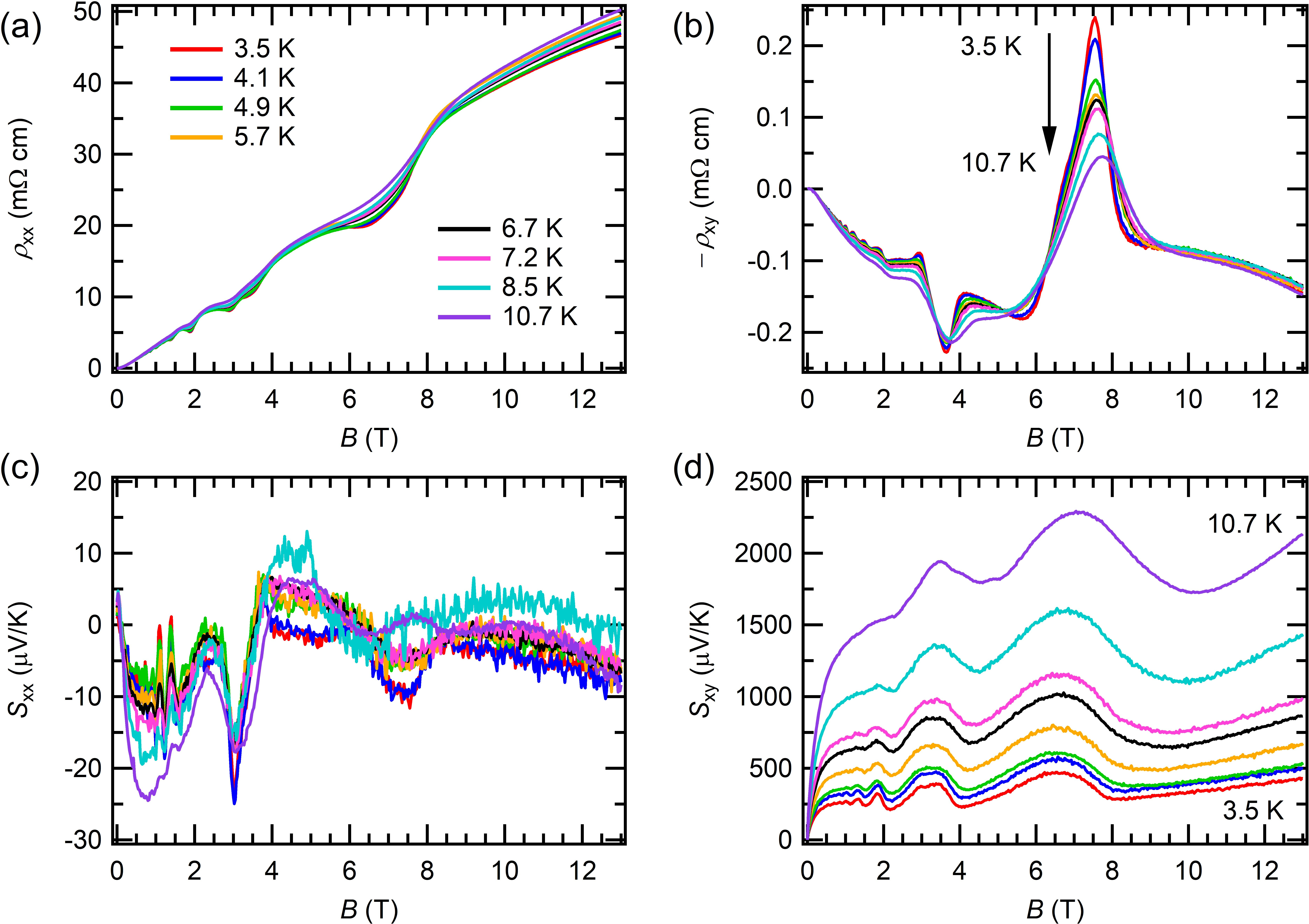}
\end{center}
\caption{(Color online) Magnetic field dependence at several fixed temperatures for (a) resistivity \rxx, (b) Hall resistivity (note: $-$\rxy), (c) Seebeck coefficient \sxx, and (d) Nernst coefficient \sxy.} \label{Fig3}
\end{figure*}

Figure \ref{Fig3} summarizes all of the measured quantities in this work. As a reference, we consider the electrical current flow and temperature gradient directions to be $\textbf{I} \parallel -\boldsymbol{\nabla} T$ and $\textbf{B} \parallel \hat{\textbf{z}}$. To standardize notation, we report the transverse coefficients as \rxy~and \sxy~unless otherwise noted. Magnetic field dependences of \rxx~and \rxy~taken at several fixed temperatures are shown in Figs. \ref{Fig3}(a) and (b), respectively [note $-\rho_{xy}$ in Fig. \ref{Fig3}(b)]. Typical Shubnikov--de Haas (SdH) oscillations are clearly seen with the last oscillation appearing at $B \approx 7.5$ T. As $B$ increases, the SdH oscillation amplitude increases, such that the last peak anomalously crosses zero. The zero crossing \rxy~has been observed in some experiments, and appears to be sample dependent with some reports show no sign change at high field \cite{Woollam, Spain, Hubbard, Jesus}. The sign change may occur in the Shubnikov--de Haas regime due to the changing balance between the number of electrons and holes, and more recently shown to be related to possible disorder upon doping. Our Kish graphite sample shows a sign change around $B = 6.8$ T, similar to the undoped sample in Refs. \cite{Jesus} and \cite{Hubbard}. Our interests here are the facts that both \rxx~and \rxy~appear to be only weakly dependent on the temperatures within the range considered and that $\left|\rho_{xy}\right| \ll \left|\rho_{xx}\right|$ throughout the field sweep.

From these data, the conductivities $\sigma_{xx}$ and $\sigma_{xy}$ can be obtained by inverting the resistivity matrix, or explicitly, $\sigma_{xx} = \rho_{xx}/ (\rho_{xx}^2 + \rho_{xy}^2)$ and $\sigma_{xy} = -\rho_{xy}/ (\rho_{xx}^2 + \rho_{xy}^2)$. Since $\left|\rho_{xy}\right| \ll \left|\rho_{xx}\right|$ one sees immediately that $\left|\sigma_{xy}\right| \ll \left|\sigma_{xx}\right|$. This is in contrast to the dissipationless limit condition $\sigma_{xx} = 0$, and is always true for graphite \cite{McClure2, Spain}. For this reason, the difference between electron and hole densities cannot be obtained by using the usual Hall coefficient $R_H \rightarrow 1/e(n-p)$, where $n$ and $p$ are the electron and the hole density, respectively \cite{Spain}. As an alternative, we used the SdH oscillations to obtain $n-p$. A Fourier transform of $d^2\rho_{xx}/dB^2$ yields two dominant components with frequencies of 4.72 and 6.42 T. These frequencies correspond to carrier pockets located near the $H$ and $K$ points, respectively, in agreement with known results \cite{Soule2, Schneider}. Taking into account the geometry and the multiplicity of the pockets in the Brillouin zone, we found $\left|n-p\right| \approx 6 \times 10^{17}$ /cm$^3$, about half of the value used in our calculations \cite{Osada}. (See also the Supplemental Material for the temperature dependence of the SdH frequencies \cite{Suppl})

Next, we look into the thermoelectric coefficients \sxx~and \sxy~shown in Figs. \ref{Fig3}(c) and (d), respectively. One again sees a typical SdH-type oscillatory behavior. In contrast to the transport coefficients where $\left|\rho_{xy}\right| \ll \left|\rho_{xx}\right|$, here \sxx~is dwarfed by \sxy~at all temperatures. Note that while \sxx~is not strongly dependent on $T$, \sxy~is very well resolved with $T$ due to its strong response at low $B$. The curves behave as \sxy $\propto T$ for $T \leq 5.7$ K throughout the magnetic field range. At $B \geq 8$ T, \sxy~gains a $B$-linear behavior. This implies $S_{xy} \propto (BT)$ in the QL regime, similar to the behavior reported in Refs. \cite{Zhu, Fauque}. At higher temperatures, the curves deviate from the $S_{xy} \propto T$ tendency, but generally remain proportional to $B$. (See also the Supplemental Material for the $B$ and $T$ dependences of \sxy~\cite{Suppl})

Having seen each of \rxx, \rxy, \sxx, and \sxy~separately, we now turn our attention to \axy. Thermoelectric Hall conductivity \axy~relates all the quantities above, and is written explicitly as
\begin{equation}
\alpha_{xy} = \frac{1}{\rho_{xx}^2 + \rho_{xy}^2} \left( \rho_{xx} S_{xy} - \rho_{xy} S_{xx}\right)
\label{eq:alphaxy1}
\end{equation}
The term \rxy\sxx~is overwhelmed by \rxx\sxy~by 3--4 orders of magnitude. Assuming a Seebeck coefficient of a gold wire of $S_{\text{Au}}\sim 1$ $\mu$V/K, its contribution will only change the overall \axy~by about 0.01\%, so neglecting its contribution in \axy~can be justified, although not necessarily negligible when considering only \sxx.

\begin{figure}[tbp]
\linespread{1}
\begin{center}
\includegraphics[width=3in]{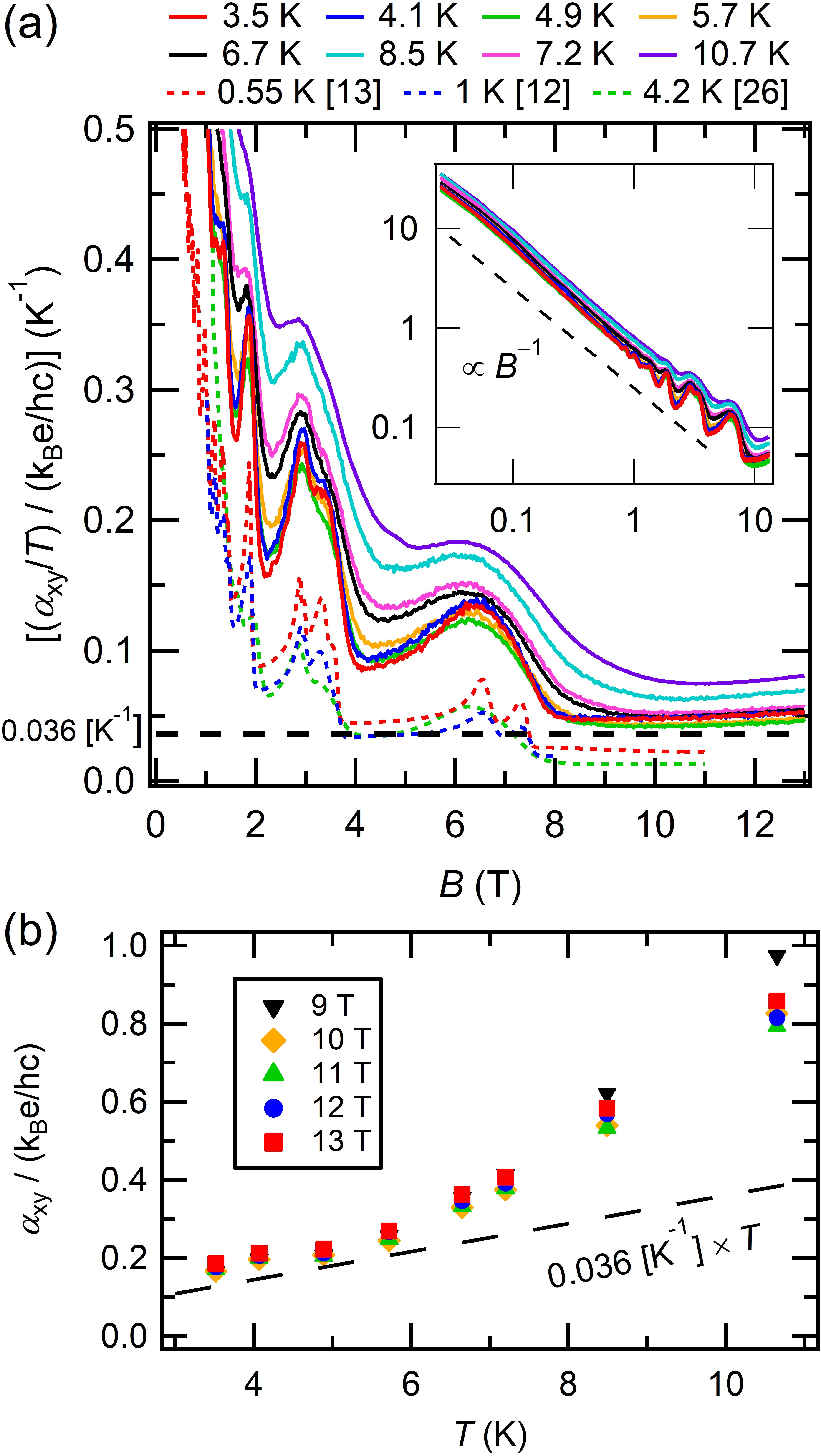}
\end{center}
\caption{(Color online) (a) Magnetic field dependence of thermoelectric Hall conductivity $(\alpha_{xy}/T) / (k_{\text{B}}e/ch)$ obtained from the transport and thermoelectric measurements showing behavior consistent with Ref. \cite{Osada}. The curves overlap with each other at $T \leq 5.7$ K and are quantized to a value that depends only on $t_c$. Dashed lines are \axy~calculated using transport data taken from Refs. \cite{Woollam}, \cite{Zhu}, and \cite{Kreps} combined with \sxy~data from Ref. \cite{Zhu}. The inset shows the low-field region $\alpha_{xy} \propto B^{-1}$ behavior consistent with the behavior shown in Fig. 1(b) for a semimetal with straight nodal lines. (b) Temperature dependence of \axy~at several fixed magnetic fields, showing a $T$-linear dependence as predicted in Eq. (2).} \label{Fig4}
\end{figure}

Considering that the coefficients other than \sxy~are not modified significantly by temperature, it is expected that the $T$-linear dependence of \sxy~is reflected in \axy. Figure \ref{Fig4}(a) shows $\alpha_{xy}/T$ calculated from the experimental data using Eq. \ref{eq:alphaxy1}. For convenience, here we plot the dimensionless $\alpha_{xy} / (k_{\text{B}}e/ch)$, where $k_{\text{B}}$ is the Boltzmann constant, $e (> 0)$ is the elementary charge, $c$ is the $c$-axis lattice constant ($c/2 =$ 0.337 nm), and $h$ is the Planck constant. The general behavior follows closely the predicted behavior for graphite illustrated in Fig. \ref{Fig1}(f).  At low fields the curves first show a monotonic decrease with $\alpha_{xy} \propto B^{-1}$ dependence as shown in the inset of Fig. \ref{Fig4}(a), consistent with the predicted behavior for the straight nodal-line semimetal shown in Fig. \ref{Fig1}(d). Up to 5.7 K, the curves follow an overall $\alpha_{xy} \propto T/B$ behavior. As $B$ is increased even further beyond the last SdH oscillation, the system enters the quasi-QL.

Upon entering the quasi-QL region \axy~changes slope and tends to a value independent of $B$. This plateau extends from the last SdH peak to the maximum field at each temperature, which is expected to be a manifestation of the QTHE. To clearly show its $T$ dependence, \axy~points taken at several fixed $B$ are shown in Fig. \ref{Fig4}(b). In a general case of multilayer semimetals with straight nodal lines, we have previously found that the plateau value is given approximately by the following, including spin and valley degeneracy \cite{Osada},
\begin{equation}
\alpha_{xy} / (k_{\text{B}}e/ch) = \frac{2 \pi k_{\text{B}} T}{3 t_c}.
\label{eq:alphaxy2}
\end{equation}
Here, $t_c$ is the interlayer transfer integral. For graphite, Eq. \ref{eq:alphaxy2} gains an additional factor of two coming from the doubly degenerate lowest LL subbands $N' = 0, -1$. The value of $t_c$ can be estimated from the width of the $B$-independent $N' = -1$ subband given by $4t_c \approx 40$ meV, and the same value was assumed for the $N' = 0$ subband. With this value, one obtains $\alpha_{xy} / (k_{\text{B}}e/ch) \approx 0.036$~(K$^{-1}$)~$\times T$ . The plateaus followed this behavior not only qualitatively, but quantitatively as well for $T \leq 5.7$ K. This fact strongly suggests that the observed behavior results from the QTHE predicted in graphite. Above 5.7 K the slope becomes steeper, indicating a deviation from $\alpha_{xy} \propto T$ although the plateau survives. This is the same deviation seen in \sxy. Theoretically, the $T$-linear behavior is expected at the low-temperature region defined by $k_{\text{B}} T \ll t_c$, which corresponds to $T \ll 110$ K for graphite. However, it should be noted that this is based on transport in the clean limit without phonon scattering. In reality, the phonon drag effect enhances both \sxx~and \sxy~\cite{Sugihara, JayGerin}. Our observation of $\alpha_{xy} \propto T$ occurs at $T \ll 15$ K, where a peak in \sxx~can be observed. This temperature is similar to the reported $S_{xx}$ peak temperature by another group \cite{Tokumoto}, and so is likely a consequence of the phonon drag effect. (See Supplemental Material for the temperature dependence of \sxx~and measurement results at extended temperatures \cite{Suppl})

Here, we compare our result with data published by other groups. We took the transport data for 0.55, 1.1, and 4.2 K from Refs. \cite{Woollam, Kreps,Zhu} and \sxy~data with the closest matching temperatures from Ref. \cite{Zhu}. As shown in Fig. \ref{Fig4}(a), \axy~calculated using these data show a plateau at QL with the correct order of magnitude, but with values lower than predicted by Eq. \ref{eq:alphaxy2}. However, differences in the reported magnitudes of $\rho_{xx}(B)$ make it difficult to compare the resulting \axy. As far as we know similar complete measurements on one sample have been performed so far only by Zhu \textit{et al}. on their sample labeled "HOPG sample 2" \cite{Zhu}. For this sample, assuming that $\rho_{xx}$ does not vary much with temperature, we approximated $\alpha_{xy} / (k_{\text{B}}e/ch) \approx 0.023$~(K$^{-1}$)~$\times T$. However, since $\alpha_{xy} \approx S_{xy}/\rho_{xx}$ and $d\rho_{xx}/dT > 0$, their $T$ dependence of $\alpha_{xy}$ likely follows a gentler slope than the estimate above.

Now, we address the slight deviation from the \axy~plateau predicted by Eq. \ref{eq:alphaxy2}. Using this equation, any deviation from the plateau can only be introduced via $t_c$, with others being some fundamental constants. In the case of graphite, however, it is not perfectly accurate to employ Eq. \ref{eq:alphaxy2} because the Fermi velocity (related to the subband width) of the $N' = 0$ LL is slightly different from that of the $N' = -1$ LL. This difference introduces a deviation of less than 1\% compared with Eq. \ref{eq:alphaxy2} at $T =$ 5 K. Additionally, whereas the Fermi velocity of the $N' = -1$ subband has no $B$ dependence, the $N' = 0$ has a weak $B$ dependence \cite{Inoue, Nakao}. The $B$ dependence of the $N'=0$ LL subband is such that its Fermi velocity decreases with increasing field, consequently \axy~tends to rise on average. This may be the reason why the QTHE plateau of \axy~deviates from $0.036$~(K$^{-1}$)~$\times (k_{\text{B}}e/ch)T$ and show a weak $B$ dependence.

Next, we comment on the nonappearance of the $B$-linear \sxx~in graphite despite the \axy~plateau. For a chiral LL of 3D DWSM, the DOS is energy-independent, which is responsible for \axy~plateau. In the dissipationless limit $(\sigma_{xx} \rightarrow 0)$, \sxx~is given by $S_{xx} \approx \alpha_{xy} / \sigma_{xy} = -\alpha_{xy} B_z / e(n - p)$. Therefore, the $B$-linear growth occurs for constant $n - p$. For graphite, the DOS is approximately energy independent, but because the system is dissipative $(\sigma_{xx} \gg \sigma_{xy})$ the \sxx~approximation above does not apply, so the $B$-linear increase of \sxx~cannot be expected. However, the \axy~plateau behavior survives because it corresponds to the dominant leading term of \axy~in the dissipative system \cite{Osada}.

Finally, the present model can be extended for systems having multiple straight nodal lines, given that those nodal lines are parallel to the applied magnetic field. Equation \ref{eq:alphaxy2} implicitly already contains a factor of two, originating from the two valleys shown in the inset of Fig. \ref{Fig1}(d). In the case of multiple straight nodal lines parallel to a magnetic field, Eq. \ref{eq:alphaxy2} is modified to include the total number of straight nodal lines. This specific case is similar to the case of DWSM with a multiple Dirac nodes discussed in Refs. \cite{Kozii} and \cite{Skinner}.

\section{IV. CONCLUSION}
In conclusion, we have demonstrated that graphite shows QTHE as a straight nodal-line semimetal. Although the system is dissipative, the dissipationless leading term of \axy~exhibiting the QTHE plateau becomes dominant. The unlimited $B$-linear increase of \sxx~cannot be expected in this case, but \axy~remains quantized due to an energy-independent density of states, similar to 3D Dirac/Weyl nodal-point semimetals. The present result shows that quantized \axy~is a strong indicator of 3D DWSM, but not their exclusive property.

\section{ACKNOWLEDGMENTS}
The authors thank Prof. Sonia Haddad for valuable discussions, and Prof. Woun Kang for his technical advice on thermoelectric measurements. This work was supported by JSPS KAKENHI Grants No. JP19K14655, No. JP20H01860, and No. JP21K18594.

\end{document}

\title{Supplemental Material: Quantized Thermoelectric Hall Plateau in the Quantum Limit of Graphite as a Nodal Line Semimetal}

\author{Andhika~Kiswandhi}
\email{kiswandhi@issp.u-tokyo.ac.jp}
\affiliation{Institute for Solid State Physics, University of Tokyo, 5-1-5 Kashiwanoha, Kashiwa, Chiba 277-8581, Japan}

\author{Tomotaka~Ochi}
\affiliation{Institute for Solid State Physics, University of Tokyo, 5-1-5 Kashiwanoha, Kashiwa, Chiba 277-8581, Japan}

\author{Toshihiro~Taen}
\affiliation{Institute for Solid State Physics, University of Tokyo, 5-1-5 Kashiwanoha, Kashiwa, Chiba 277-8581, Japan}

\author{Mitsuyuki~Sato}
\affiliation{Institute for Solid State Physics, University of Tokyo, 5-1-5 Kashiwanoha, Kashiwa, Chiba 277-8581, Japan}

\author{Kazuhito~Uchida}
\affiliation{Institute for Solid State Physics, University of Tokyo, 5-1-5 Kashiwanoha, Kashiwa, Chiba 277-8581, Japan}

\author{Toshihito Osada}
\email{osada@issp.u-tokyo.ac.jp}
\affiliation{Institute for Solid State Physics, University of Tokyo, 5-1-5 Kashiwanoha, Kashiwa, Chiba 277-8581, Japan}
\date{\today}% It is always \today, today,

\maketitle

\section{S1. Sample preparation and configuration}\label{sectionS1}

%\begin{figure*}[tbp]
%\linespread{1}
%\begin{center}
%\includegraphics[width=5in]{figureS1.jpg}
%\end{center}
%\caption{(left) Photograph of the experimental setup and (right) schematic of the experimental setup corresponding to the photograph shown on the left panel. The sample is attached between a heater and copper heat sink. Two thermocouples (broken line) measure the temperatures on the heater side and heat sink side of the sample. The thermocouples were thermally anchored to the heat sink adjacent to a Cernox thermometer. The symbols denote electrode connections to the sample, with ($+$) and ($-$) superscripts denote positive and negative electrodes, respectively.}\label{FigS1}
%\end{figure*}

In this study, we performed \rxx, \rxy, \sxx, and \sxy~measurements all on a single sample. The sample was selected due to its size and quality (for residual resistivity ratio, see section S2). The sample was cut into approximate rectangular shape and cleaved using a weak adhesive tape then washed with acetone to remove any residual adhesive. On the clean side, six gold wires (25 $\mu$m) were attached as electrodes using silver paste. The sample was then attached using Stycast 2850 FTJ epoxy to a platform lengthwise with one side on a chip resistor heater and the opposite side on a copper heat sink. The final configuration is shown in Fig. 2 of the main article.

For the electrical transport measurement, a Cernox thermometer, attached to the heat sink, was used to monitor the temperature. For the thermoelectric measurement, we used one heater two thermometers setup. Here, in addition to the Cernox, two fine wire (25 $\mu$m) chromel-constantan thermocouples were used. The junction of each thermocouple was lightly coated with Stycast 2850FTJ epoxy to prevent short circuit to the sample. The thermocouples were referenced to the heat sink and attached onto the sample near its ends using the same epoxy. With this setup, the temperatures at sample ends can be determined as
\begin{equation}
T_{\text{h(c)}}=T_{\text{sink}}+\frac{V_{\text{h(c)}}}{S_{\text{TC}}(T_{\text{sink}})}
\label{eq:STC}
\end{equation}
where $T_{\text{sink}}$ is the temperature at the heat sink measured by the Cernox thermometer, $S_{\text{TC}}(T_{\text{sink}})$ is the Seebeck coefficient of the thermocouple at the heat sink temperature, $T_{\text{h(c)}}$ and $V_{\text{h(c)}}$ are the temperature and the thermocouple voltage, respectively measured at the hot (cold) end. The sample temperatures in the thermoelectric measurements were determined as the average between the two temperatures.

All measurements were carried out with dc mode using Keithley 220 current source and Keithley 2182 nanovoltmeter. For the electrical transport measurement, current reversal technique was used. To remove contamination between the diagonal ($xx$) and off-diagonal ($xy$) components due to contact misalignment, standard symmetrization and antisymmetrization procedures were used. Therefore, the reported magnetic field dependent \rxx~and \rxy~were obtained as. 
\begin{equation}
\begin{aligned}
\rho_{xx}(B) = \frac{\rho_{xx}(B\geq0) + \rho_{xx}(B\leq0)}{2} \\
\rho_{xy}(B) = \frac{\rho_{xy}(B\geq0) - \rho_{xy}(B\leq0)}{2}.
\label{eq:rhosymm}
\end{aligned}
\end{equation}
As for the dc thermoelectric measurements, the signal was obtained after subtracting the voltage measured in the heater off state from the voltage in the heater on state to minimize the zero offset. Similar to the resistivity data, \sxx~and \sxy~were symmetrized and antisymmetrized to eliminate contact offset error as the following
\begin{equation}
\begin{aligned}
S_{xx}(B) = \frac{S_{xx}(B\geq0) + S_{xx}(B\leq0)}{2} \\
S_{xy}(B) = \frac{S_{xy}(B\geq0) - S_{xy}(B\leq0)}{2}.
\label{eq:rhosymm}
\end{aligned}
\end{equation}

Since gold wires were used as electrodes, it is important to note that the \sxx~values are measured relative to gold such that $S_{xx} = S_{xx}^{\text{sample}}-S_{\text{Au}}$. At low temperatures, the value of $S_{\text{Au}}$ is in the order of 1 $\mu$V/K even under magnetic field up to 30 T \cite{Choi1, Choi2}.  As shown in the discussion part, for our purposes the contribution from the term containing \sxx~in \axy~is small even after including $S_{\text{Au}}$. However, it should be understood not to deduce the carrier type using \sxx~directly since $S_{\text{Au}}$ may not be negligible when considering only \sxx, so \sxx~is not the same as the absolute $S_{xx}^{\text{sample}}$.

\section{S2. Residual resistivity ratio and sample quality}\label{sectionS2}

\begin{figure}[htbp]
\linespread{1}
\begin{center}
\includegraphics[width=6.25in]{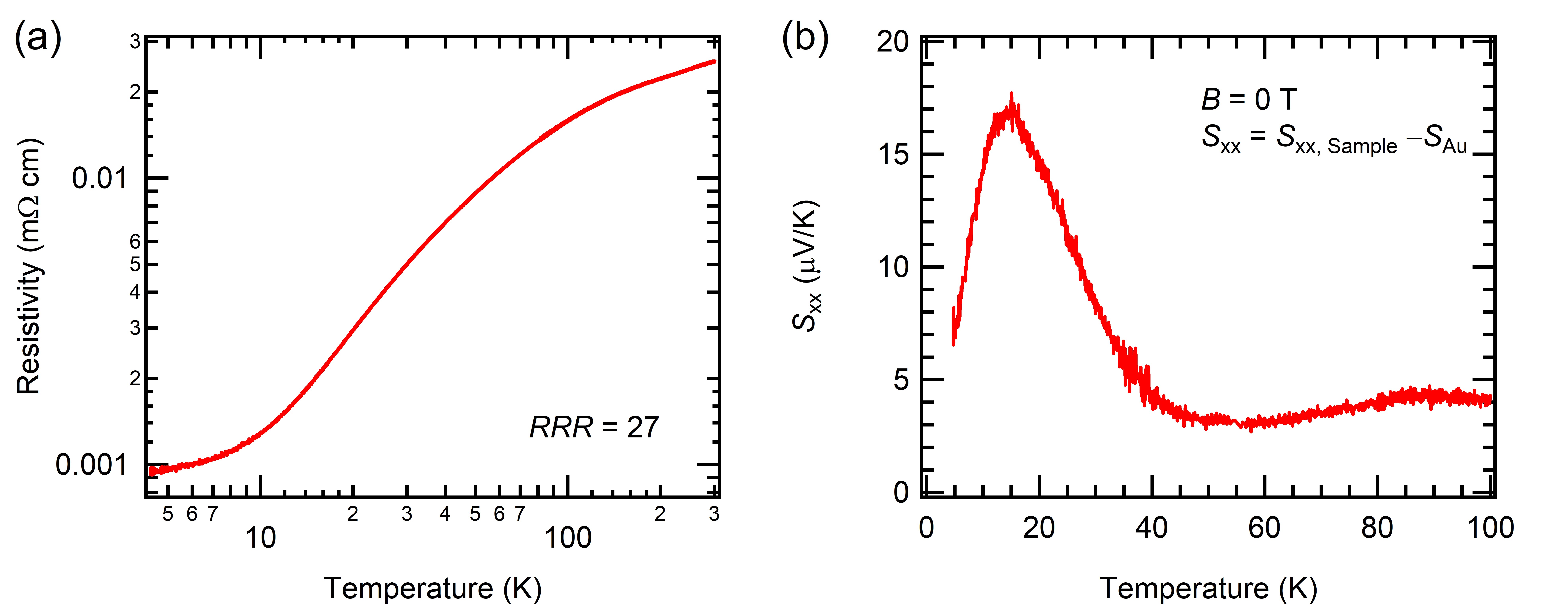}
\end{center}
\caption{Temperature dependence of (a) resistivity $\rho_{xx}$ and (b) Seebeck coefficient $S_{xx}$ obtained at zero magnetic field.}\label{FigS1}
\end{figure}

Natural graphite has a wide range of quality, which may affect its electronic properties \cite{Kaburagi}. We checked the sample quality by obtaining residual resistivity ratio $RRR = R(\text{300 K}) / R(\text{4.2 K})$ from temperature dependent resistivity data at zero field shown in Fig. \ref{FigS1}(a). The sample shows $RRR =$ 27, as compared with typical $RRR >$ 10 for natural graphite \cite{Soule1}.

\section{S3. Temperature dependence of Seebeck Coefficient \sxx}\label{sectionS3}

In the main text, we provide data up to $T = 10.7$ K. From Fig. 4 it can be seen that deviation from $\alpha_{xy} \propto T$ can already be seen above $T = 5.7$ K, above which \axy~increases with a steeper slope. From Fig. 3, we notice that the temperature dependence of \axy~should follow the temperature dependence of \sxy~because $\alpha_{xy} \approx S_{xy}/\rho_{xx}$ and other coefficients are almost temperature independent. It has been reported that phonon drag gives a significant enhancements in both \sxx~and \sxy~in graphite \cite{Sugihara, JayGerin}. Since $\alpha_{xy} \propto T$ behavior is based on the clean limit, the behavior can only be expected at temperatures where phonon scattering effect is relatively weak. This means that QTHE is expected at $T \ll 15$ K corresponding to the \sxx~peak seen in Fig. \ref{FigS1}(b), which is similar to the temperature reported in ref. \cite{Tokumoto}. 

\section{S4. Temperature dependence of carrier concentration}\label{sectionS4}

\begin{figure*}[htbp]
\linespread{1}
\begin{center}
\includegraphics[width=6.25in]{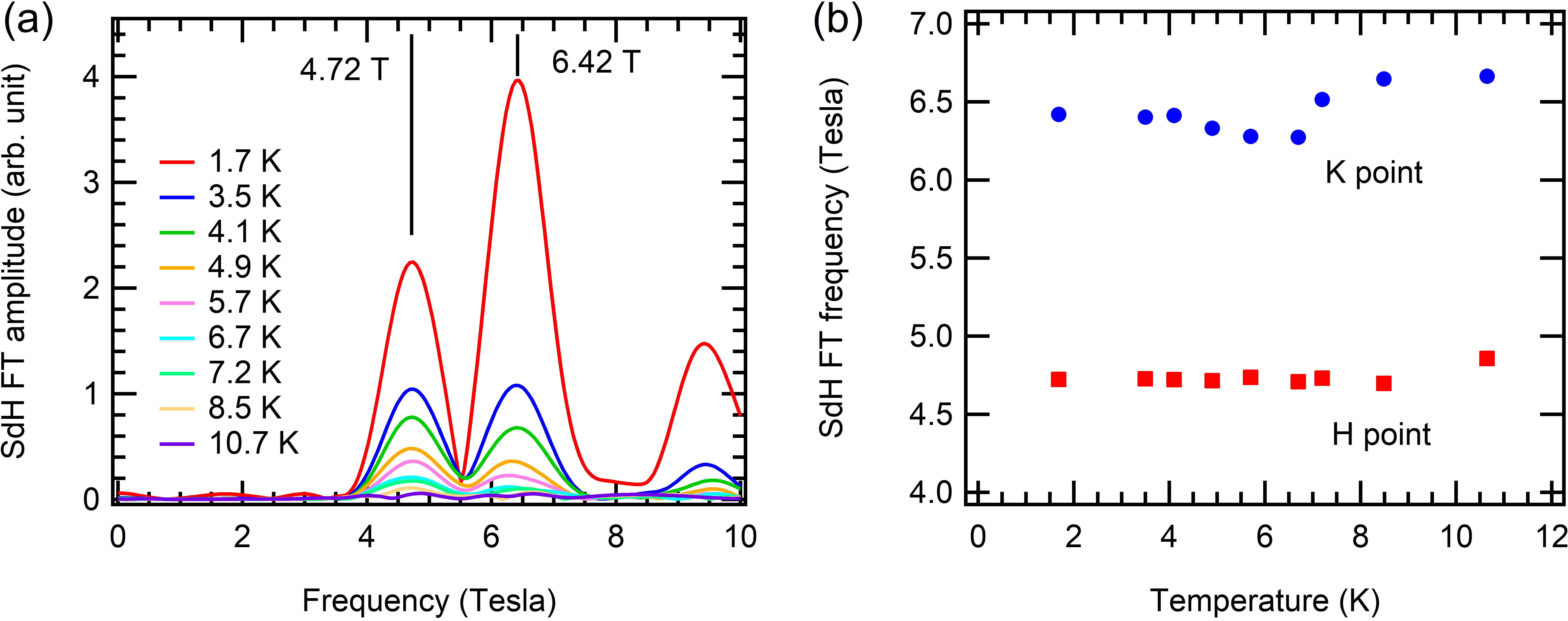}
\end{center}
\caption{(a) SdH peaks obtained from Fourier transform of $d^2\rho_{xx}/dB^2$ data at various temperatures. Dominant peaks appear about 4.72 T and 6.42 T, with an additional harmonic appearing around 9.4 T. (b) temperature dependence of SdH frequency.}\label{FigS2} 
\end{figure*}

Since graphite is dissipative due to $\left|\rho_{xy}\right| \ll \left|\rho_{xx}\right|$, its carrier density difference $n - p$ should not be obtained from its Hall coefficient $R_H \rightarrow 1/e(n-p)$. Here $n$ and $p$ denote the electron and the hole density, respectively. Alternatively, the Hall conductivity formula $\sigma_{xy} \rightarrow e(n-p)/B$ may be used \cite{Spain}. However, this approximation appears to underestimate the carrier density difference, for example $n - p \sim 10^{16}$/cm$^3$ in Ref. \cite{Spain}. We also found the same order of magnitude $n - p$ for our sample using the above approximation. To avoid this ambiguity, we determine $n - p$ from Shubnikov-de Haas (SdH) oscillations by taking Fourier transform of the second derivative $d^2\rho_{xx}/dB^2$ shown in Fig. \ref{FigS2}. Taking into account the multiplicity of carrier pockets in the Brillouin zone (4 for H-point and 2 for K-point) and ellipsoidal Fermi surface (major axis : minor axis = 12 : 1) \cite{Soule2}, we obtained $\left| n - p \right| \approx 6\times 10^{17}$/cm$^3$.

\section{S5. Magnetic field and temperature dependence of \sxy}\label{sectionS5}

\begin{figure*}[htbp]
\linespread{1}
\begin{center}
\includegraphics[width=6.25in]{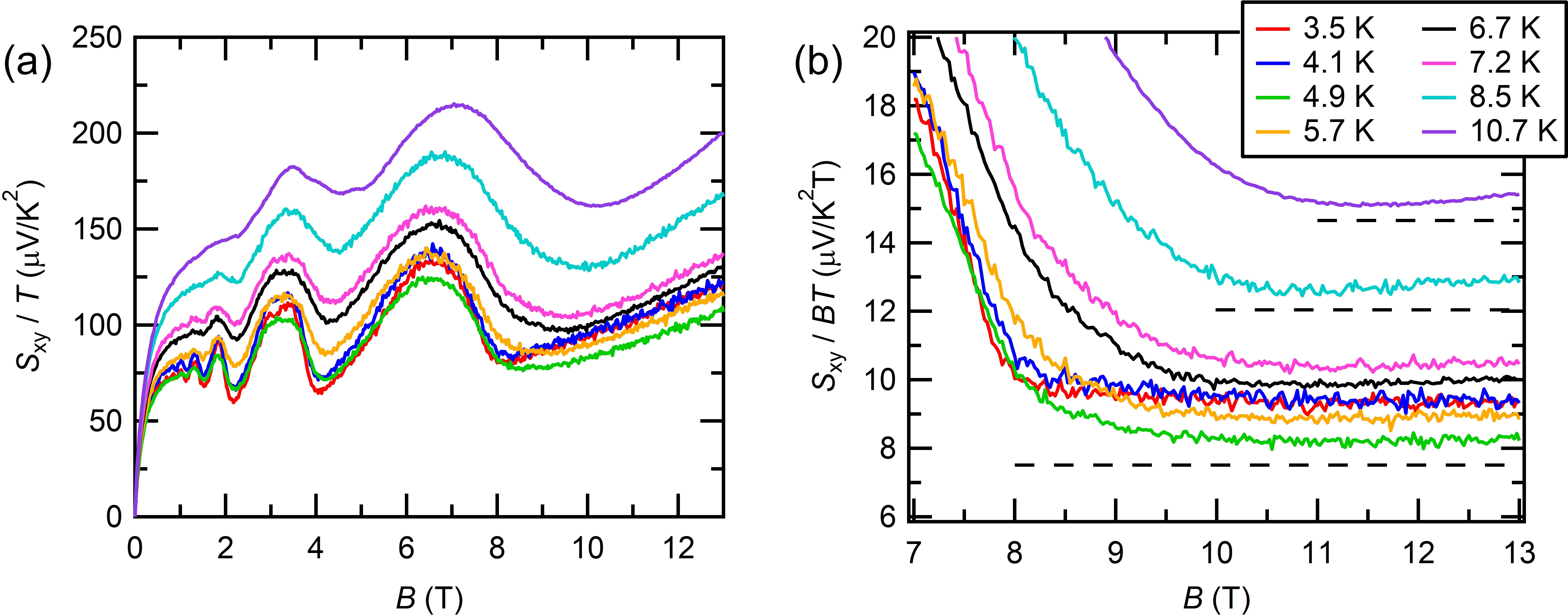}
\end{center}
\caption{(a) Magnetic field dependence of $S_{xy}/T$ for various temperatures. (b) Magnetic field dependence of $S_{xy}/(BT)$ in quasi-quantum limit region taken at various temperatures.}\label{FigS3}
\end{figure*}

The Nernst coefficient \sxy~in graphite has a large magnetic field response and well resolved with temperature. The dependence is reflected in thermoelectric Hall conductivity \axy~data. In this section we describe the \sxy~dependence on $B$ and $T$ in more detail.

First, we investigate the temperature dependence of \sxy. For this purpose, we analyze $S_{xy}/T$ shown in Fig. \ref{FigS3}(a). The curves appear to converge for low temperature range $T \leq 5.7$ K for the whole magnetic field range. This indicates $S_{xy} \propto T$ tendency in this region, which has been reported by Zhu \textit{et al}. \cite{Zhu}. When the temperature is increased, the curves deviate from the $S_{xy} \propto T$ tendency as evidenced by increasing $S_{xy}/T$ as $T$ increases.

We then look into the quasi-quantum limit (quasi QL) part at $B > 7.5$ T. From Fig. 3(d) in the main text and Fig. \ref{FigS3}(a), one can see $S_{xy}(B)$ appears proportional to $B$. Therefore, we take $S_{xy}/(BT)$ as shown in Fig. \ref{FigS3}(b). The resulting curves appear to be $B$-independent, which indicate an additional $S_{xy} \propto B$ tendency at high field region. Interestingly, as temperature increases this behavior seems to persist, although deviation can be seen appearing about $T > 8.5$ K. With this, we conclude that in the QL region \sxy~follows $S_{xy} \propto (BT)$ dependence, as mentioned in the main text, confirming the reported behavior in Ref. \cite{Zhu}.

\section{S6. Transport, thermoelectric coefficients, and thermoelectric Hall Conductivity at High Temperatures}
\begin{figure*}[htbp]
\linespread{1}
\begin{center}
\includegraphics[width=5.4in]{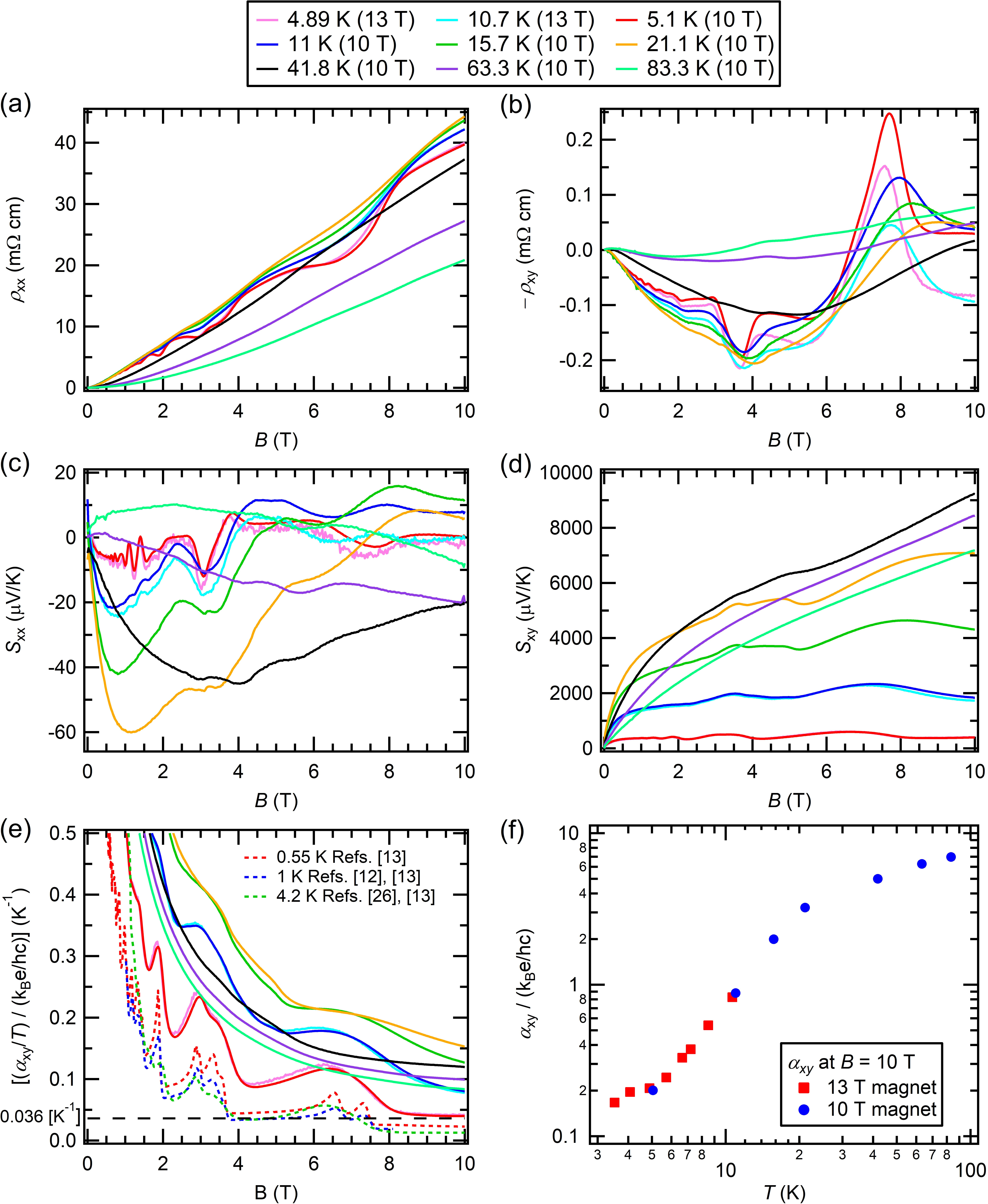}
\end{center}
\caption{Magnetic field dependence of (a) $\rho_{xx}$, (b) $\rho_{xy}$, (c) $S_{xx}$, (d) $S_{xy}$, and (e) $(\alpha_{xy}/T) / (k_{\text{B}}e/ch)$ for various temperatures. Data labeled as 4.89 K (13 T) and 10.7 K (13 T) are those taken using 13 T magnet system for reference, while the rest were taken using 10 T magnet system. In (e), $(\alpha_{xy}/T) / (k_{\text{B}}e/ch)$ calculated using data taken from Refs. [12], [13], and [26] of the main text are plotted for comparison. (f) Temperature dependence of $(\alpha_{xy}/T) / (k_{\text{B}}e/ch)$ at $B = 10$ T. Red symbols are points taken from measurement using 13 T magnet system, while blue circles are from 10 T magnet system.}\label{FigS4}
\end{figure*}
Thermoelectric Hall conductivity at temperatures where QTHE can be observed and where deviation from QTHE due to phonon drag effect starts becoming significant has been discussed. In this section, we provide data at extended temperatures up to 83.3 K on the same sample. For this measurement, however, we used a different magnet system with a maximum field of 10 T.

Since the extended temperature data were taken separately with a different magnet system, we first compared the results taken using the 10 T at 5.1 and 11 K with the results from the 13 T magnet systems at 4.89 K and 10.7 K. As shown in Fig. \ref{FigS4}, the measurement results from the 10 T and 13 T agree with each other. There are some differences notably in the high field part of \rxy~and \sxx, but the resulting \axy~are in agreement with each other. Again, this is because \axy~is dominated by $\rho_{xx} S_{xy}$ term, which is 3 - 4 orders of magnitude larger than $\rho_{xy} S_{xx}$ term. 

At near zero field, \sxx~and \sxy~both show firstly a monotonic increase in magnitude up to 21.1 K, followed by a decrease as the temperature is increased further. This behavior appears consistent with the reported phonon drag effect, which enhances both \sxx~and \sxy~\cite{Sugihara, JayGerin, Tokumoto}. It can be seen in the $\alpha_{xy}/T$ data in Fig. \ref{FigS4}(e) and (f) that with increasing temperature the deviation from $\alpha_{xy} \propto T$ increases and peaked at 21.1 K, then decreases with further temperature increase. At high temperature, however, it will be difficult to reach the quantum limit, so \axy~cannot be expected to reach the QTHE plateau.